\documentclass[twocolumn,floats,prd,amssymb,floatfix,nofootinbib,balancelastpage,superscriptaddress,amsmath]{revtex4-1}
\pdfoutput=1
\usepackage[utf8]{inputenc}
\usepackage{amssymb}
\usepackage{amsmath}
\usepackage{amsfonts}
\usepackage{graphicx}
\usepackage{color}
\usepackage{xspace}

 \usepackage[section]{placeins}
\usepackage{afterpage}

\usepackage{float}
\usepackage{slashed}

\newcommand{\CODEXb}{\operatorname{\mathtt{CODEX-b}}}

\usepackage{multirow,rotating}

\def\gsim{\raise0.3ex\hbox{$\;>$\kern-0.75em\raise-1.1ex\hbox{$\sim\;$}}}
\def\lsim{\raise0.3ex\hbox{$\;<$\kern-0.75em\raise-1.1ex\hbox{$\sim\;$}}}

\newcommand{\ba}[1]{\begin{eqnarray} \label{(#1)}}
\newcommand{\ea}{\end{eqnarray}}

\newcommand{\lam}{\lambda}

\definecolor{tobycolour}{rgb}{.6,.0,.4}

\definecolor{jcolour}{rgb}{.6,.5,.4}

\definecolor{dcolour}{rgb}{.5, .5, .5}

\def\gsim{\raise0.3ex\hbox{$\;>$\kern-0.75em\raise-1.1ex\hbox{$\sim\;$}}}
\def\lsim{\raise0.3ex\hbox{$\;<$\kern-0.75em\raise-1.1ex\hbox{$\sim\;$}}}

\begin{document}


\title{R-parity Violation and Light Neutralinos at CODEX-b, FASER, and MATHUSLA}

\author{Daniel Dercks}
\email{daniel.dercks@desy.de}
\affiliation{II. Institut f\"ur Theoretische Physik, Universit\"at Hamburg,\\ Luruper Chaussee 149, 22761 Hamburg, Germany}

\author{Jordy de Vries}
\email{j.de.vries@fz-juelich.de}
\affiliation{Amherst Center for Fundamental Interactions, Department of Physics, \\ University of Massachusetts, Amherst, MA 01003, USA}
\affiliation{RIKEN BNL Research Center, Brookhaven National Laboratory, Upton, NY 11973-5000, USA}

\author{Herbi K. Dreiner}
\email{dreiner@uni--bonn.de}
\affiliation{Physikalisches Institut der Universit\"at Bonn, Bethe Center
for Theoretical Physics, \\ Nu{\ss}allee 12, 53115 Bonn, Germany}

\author{Zeren Simon Wang}
\email{wzeren@physik.uni-bonn.de}
\affiliation{Physikalisches Institut der Universit\"at Bonn, Bethe Center
for Theoretical Physics, \\ Nu{\ss}allee 12, 53115 Bonn, Germany}





 \begin{abstract}
The $LQ\bar D$ operator in R-parity-violating supersymmetry can lead to meson decays to light neutralinos and 
 neutralino decays to lighter mesons, with a long lifetime. Since the high-luminosity LHC is expected to accumulate 
 as much as 3/ab of data, several detectors proposed to be built at the LHC may probe unexplored regions in the parameter 
 space, for long-lived neutralinos. We estimate the sensitivity of the recently proposed detectors, 
 CODEX-b, FASER, and MATHUSLA, for detecting such light neutralinos singly produced from $D$- and $B$-meson 
 decays in a list of benchmark scenarios, and discuss the advantages and disadvantages of the proposed detectors 
 in this context. We also present our results in a model independent fashion, which can be applied to any long-lived
 particle with mass in the GeV regime.
\end{abstract}


\vskip10mm

\maketitle
\flushbottom
%
%
\section{Introduction}
\label{sect:intro}
The discovery of a Standard-Model (SM) like Higgs boson in 2012 has been a highlight of the Large Hadron Collider (LHC) 
\cite{Aad:2012tfa,Chatrchyan:2012xdj}. The small Higgs mass, $m_h=125.09$ GeV \cite{Patrignani:2016xqp}, has since, 
however, consolidated the hierarchy problem \cite{Gildener:1976ai,Veltman:1980mj}. Supersymmetric (SUSY) theories offer 
an elegant solution, for reviews see Ref.~\cite{Nilles:1983ge,Martin:1997ns}. All searches for the new fields predicted by 
SUSY however, have been unsuccessful yet. This leads to lower limits on the masses of squarks and gluinos in various 
supersymmetric models of order 
$1\,$TeV and above \cite{Aad:2015wqa, Khachatryan:2015vra,Tanabashi:2018oca,Bechtle:2015nua,Dercks:2017lfq}. On the 
other hand, the lightest neutralino is not similarly constrained. In fact, if we drop the assumption for the gaugino masses,
$M_1=\frac{5}{3}\tan^2{\theta_W}M_2$, which is motivated by grand unified theories (GUTs), and do not require
the lightest neutralino to comprise the dark matter of the Universe, light neutralino masses below $\sim 10\,$GeV are still allowed 
\cite{Choudhury:1999tn,Hooper:2002nq,Bottino:2002ry,Dreiner:2009ic,Vasquez:2010ru,Calibbi:2013poa}. For neutralino massess 
between about 1\,eV and 10\,GeV the relic energy density of the neutralinos would overclose the Universe \cite{Bechtle:2015nua},
thus such neutralinos must decay. Since they are light they will typically have long lifetimes.

In R-parity-violating (RPV) SUSY (for reviews see Ref.~\cite{Barbier:2004ez,Dreiner:1997uz,Mohapatra:2015fua}), the lightest neutralino 
is no longer stable and decays via RPV couplings.\footnote{Incidentally in such RPV SUSY models the dark matter can be composed 
of axinos \cite{Chun:1999cq,Kim:2001sh,Colucci:2015rsa,Colucci:2018yaq}.} Thus such a neutralino can be light. For small couplings and 
small mass these neutralinos can be long-lived enough to escape the reach of the LHC detectors. Moreover, the RPV couplings can induce 
single production of neutralinos via rare meson decays. Such scenarios have been investigated in various fixed-target set-ups 
\cite{Choudhury:1999tn,Dedes:2001zia,Dreiner:2002xg,Dreiner:2009er}. More recently they have also been studied in the context of
the proposed  \texttt{SHiP} experiment \cite{Alekhin:2015byh,Gorbunov:2015mba,deVries:2015mfw}. Ref.~\cite{deVries:2015mfw} 
studied the expected LHC sensitivity to such scenarios assuming an integrated luminosity of 250/fb. Ignoring differences in the 
reconstruction efficiency the sensitivity in the R-parity violating couplings at \texttt{ATLAS} was lower than at \texttt{SHiP} by 
roughly a factor of 2.

It is expected that the high-luminosity LHC (HL-LHC) will deliver up to $3$/ab of luminosity in the coming 20 years 
\cite{High-Luminosity:2114693}. As the cross sections for producing long-lived particles (LLPs) are typically small, such a large 
amount of data is required for high sensitivity to LLPs. Unsurprisingly, there have appeared several proposals to build new detectors 
near the interaction points (IPs) at the LHC, exploiting the projected large luminosity: \texttt{CODEX-b} \cite{Gligorov:2017nwh}, 
\texttt{FASER} \cite{Feng:2017uoz} and \texttt{MATHUSLA} \cite{Chou:2016lxi}. In this study, we estimate the sensitivity reach of 
these detectors for discovering singly produced light neutralinos from $D$- and $B$-mesons via RPV $LQ\bar D$ couplings, and compare 
them with each other. We also interpret our studies in a model independent way, independently of the RPV couplings. Instead we 
set bounds on the product of the branching ratios of the production of an LLP from a meson decay and the decay of the LLP to a 
meson and charged lepton  in terms of the neutralino decay length $c\tau$. This can be applied to any potential LLP.

\texttt{CODEX-b} is a comparatively small cubic detector making use of a shielded space near the \texttt{LHCb} 
IP that is expected to be free soon. Since it is to be installed at \texttt{LHCb} instead of \texttt{ATLAS} or \texttt{CMS}, \texttt{CODEX-b} 
will have an expected luminosity of $300/$fb, one order of magnitude smaller than \texttt{ATLAS} or \texttt{CMS}, if \texttt{LHCb} 
runs until 2035 with upgrades to a Phase-II \cite{Aaij:2244311}. \texttt{FASER} was proposed as a small cylindrical detector to be built 
in the very forward region several hundred meters downstream of the \texttt{ATLAS} or \texttt{CMS} IP. In comparison, 
\texttt{MATHUSLA} would be built as a massive surface detector above the \texttt{ATLAS} IP. The details of the experimental setups are
summarized in Sec.~\ref{sect:exp}.

The \texttt{CODEX-b} physics proposal \cite{Gligorov:2017nwh} examined two benchmark models, \textit{i.e.} Higgs decay to dark photons, 
and $B$-meson decays via a Higgs mixing portal. Several \texttt{FASER} papers have respectively studied dark photons produced through 
light meson decays and photon bremsstrahlung \cite{Feng:2017uoz}, dark Higgs bosons produced through $B$- and $K$-mesons 
\cite{Feng:2017vli}, heavy neutral leptons \cite{Kling:2018wct} and axion-like particles \cite{Feng:2018pew}. There are also studies 
investigating \texttt{MATHUSLA} with dark Higgs \cite{Evans:2017lvd}, exotic Higgs decays to LLPs \cite{Chou:2016lxi,Curtin:2017izq}, and 
the Dynamical Dark Matter framework \cite{Curtin:2018ees}. Recently a \texttt{MATHUSLA} white paper \cite{Curtin:2018mvb} appeared, 
where the theory community presented detailed studies of \texttt{MATHUSLA}'s potential of detecting LLPs in many different models. 
Ref.~\cite{Helo:2018qej} studied all these three detectors with heavy neutral leptons in the Type-I Seesaw model, and the lightest 
neutralino pair-produced from Z bosons with the RPV-SUSY model. Very recently Ref.~\cite{Berlin:2018jbm} investigated inelastic dark matter 
models at various existing and proposed LHC experiments including \texttt{CODEX-b}, \texttt{FASER} and \texttt{MATHUSLA}. We extend this 
work to consider the production of supersymmetric neutralinos via both $D$- and $B$-mesons, as well as the decays of the neutralinos to a 
charged meson and a charged lepton. RPV-SUSY is a complete model and we thus also consider the full kinematic constraints due to phase 
space. The mass differences between the mesons, the neutralino and a potential tau-lepton strongly affect the search sensitivities.

It is the purpose of this paper to investigate the discovery potential of light neutralinos at the detectors \texttt{CODEX-b}, \texttt{FASER}, and 
\texttt{MATHUSLA}. The primary motivation of this scenario is that supersymmetry is a potential solution to the hierarchy problem.
Light neutralinos are consistent with all laboratory and astrophysical data \cite{Dreiner:2003wh,Dreiner:2009ic,Dreiner:2013tja,Helo:2018qej}.
Thus this is an allowed supersymmetric parameter range, and should be investigated. Such a light neutralino is only consistent with the
observed dark matter density if it decays on time-scales much shorter than the age of the universe. This is the case for R-parity violating 
scenarios. R-parity violating supersymmetry naturally obtains light neutrino masses, without introducing a super heavy see-saw Majorana mass
of order $10^{10}\,$GeV or higher \cite{Hall:1983id,Davidson:2000uc}. The scenario of an $\mathcal{O}(1\,\mathrm{GeV})$ neutralino
does not in itself resolve any discrepancy between the Standard Model and current data.

This paper is organized as follows. In Sec.~\ref{setc:theory} we briefly introduce the model of RPV-SUSY. In Sec.~\ref{sect:exp} we 
describe the three experiments for which we estimate the sensitivities, and explain the details of the numerical simulation. In Sec.~\ref{sect:results} 
we present results for various benchmark choices for RPV couplings. 
We summarize and conclude in Sec.~\ref{sect:conclusion}.

\section{Supersymmetry with RPV, Production and Decay of Light Neutralinos}
\label{setc:theory}

We give a brief introduction to the RPV-SUSY model, and describe the production and decay of light 
neutralinos via RPV couplings. Compared to the R-parity conserving (RPC) supersymmetric theories, RPV-SUSY 
has extra terms in the superpotential:\footnote{For a discussion 
of baryon- and lepton-number violating non-holomorphic terms in the K\"ahler pootential see 
Refs.~\cite{Csaki:2013jza,Csaki:2015fea,Csaki:2015uza}.}
\begin{align}
W_{\text{RPV}}= &\kappa_i L_i H_u + \lambda_{ijk}L_i L_j E^c_k + \nonumber \\
&\lambda'_{ijk}L_i Q_j D^c_k + \lambda''_{ijk} U_i^c D_j^c D_k^c,
\end{align}
where the first three terms are lepton number violating (LNV) and the last is baryon number violating (BNV). The 
co-existence of LNV and BNV terms would lead to too fast proton decays, so in our study we choose to be exclusively 
interested in the $LQ\bar D$ operators. With non-vanishing RPV couplings, the lightest supersymmetric particle (LSP) 
is not stable and can decay to SM particles. If the lightest neutralino is sufficiently light, it can be the LSP. We assume 
this is the case in our study.

Neutralinos that are produced from charm and bottom meson decays are necessarily lighter than $10$ GeV and are 
dominantly bino-like to avoid existing bounds, see Ref.~\cite{Dreiner:2009ic}. Formulas for the partial widths of heavy 
meson decays and for the partial widths of neutralino decays via 
$LQ\bar D$ couplings can be found in Refs.~\cite{Dedes:2001zia,Choudhury:1999tn,Dreiner:2009ic,deVries:2015mfw}.

In principle, one single $L Q \bar D$ coupling introduces several effective SM operators and hence may simultaneously 
induce both meson decays to neutralinos and neutralino decays to lighter mesons. However, as Ref.~\cite{deVries:2015mfw} 
points out, due to kinematic constraints only the coupling $\lambda'_{112}$ may lead to such a complete decay chain:
\begin{eqnarray}
K^0_{L/S}\rightarrow \tilde{\chi}_1^0\nu,~\tilde{\chi}_1^0\rightarrow K^{\pm}l^{\mp}.
\end{eqnarray}
Moreover, since the mass difference between $K^0_{L/S}$ and $K^{\pm}$ is only $4$ MeV, the kinematically allowed 
neutralino mass range is very small and  this case is not worth studying. Therefore, we only consider scenarios with two 
distinct non-vanishing RPV operators, one for the production of the neutralinos and the other for the decay.

The couplings $\lambda'_{ijk}$ for the operator $L_iQ_j\bar D_k$ have strict bounds from different sources, though the 
bounds can be substantially weakened for heavy sfermion masses above 1 TeV. For reviews, see 
Refs.~\cite{Allanach:1999ic,Barger:1989rk, Bhattacharyya:1997vv, Barbier:2004ez,Kao:2009fg}. Since we investigate 
the same benchmark scenarios as  in Ref.~\cite{deVries:2015mfw}, we only list the relevant bounds, reproduced from 
Ref.~\cite{Kao:2009fg}:
\begin{eqnarray}
\lambda'_{112} &< 0.03\;\displaystyle\frac{m_{\tilde s_R}}{100\,\mathrm{GeV}},\qquad\lambda'_{121} &< 0.2\;\displaystyle
\frac{m_{\tilde d_R}}{100\,\mathrm{GeV}},
\label{eq:single_boundsa}
\\[3mm]
\lambda'_{122} &< 0.2\;\displaystyle\frac{m_{\tilde s_R}}{100\,\mathrm{GeV}},\qquad\lambda'_{131} &< 0.03\;\displaystyle
\frac{m_{\tilde t_L}}{100\,\mathrm{GeV}},
\label{eq:single_boundsb}
\\[3mm]
\lambda'_{312} &< 0.06\;\displaystyle\frac{m_{\tilde s_R}}{100\,\mathrm{GeV}},\qquad \lambda'_{313} &< 0.06\;\displaystyle
\frac{m_{\tilde b_R}}{100\,\mathrm{GeV}}.
\label{eq:single_boundsc}
\end{eqnarray}
Some pairs of operators have even stricter product bounds. We take the relevant bounds from Ref.~\cite{Allanach:1999ic}:

\begin{eqnarray}
\sqrt{\lambda'_{121}\lambda'_{112}} &< 3\times 10^{-5}\;\displaystyle\frac{m_{\tilde \nu_L}}{100\,\mathrm{GeV}},\label{eq:product_boundsa}
\\[3mm]
\sqrt{\lambda'_{122}\lambda'_{112}} &< 4.7\times 10^{-3}\;\displaystyle\frac{m_{\tilde s_R}}{100\,\mathrm{GeV}},\label{eq:product_boundsb}
\\[3mm]
\sqrt{\lambda'_{131}\lambda'_{112}} &< 4.7\times 10^{-3}\;\displaystyle\frac{m_{\tilde e_L}}{100\,\mathrm{GeV}}.
\label{eq:product_boundsc}
\end{eqnarray}

Throughout this work, we assume that all sfermions have degenerate masses $m_{\tilde f}$. This allows us to directly 
compare the above bounds to our results even though the respectively relevant operators depend on the masses of 
possibly different SUSY particles. Note that results for significantly non-degenerate SUSY spectra may therefore differ 
significantly and can change the relative importance of bounds from different sources.

\section{Experimental Setups and Simulation} 
\label{sect:exp}

In this section we summarize the setups of the detectors, and explain in detail our simulation procedure. 
For more information on the proposed detectors we refer to Refs.~\cite{Gligorov:2017nwh,Feng:2017uoz,Chou:2016lxi}. 
The main difference between them lies in the projected luminosity, the geometry and installed position (large or 
small pseudorapidity $\eta$). Moreover, whether installed underground or above the ground, these proposals 
all argue that the background influence, for example from cosmic rays, can be well controlled. Therefore, we do 
not discuss it and always assume $100\%$ detector efficiency.

In order to estimate the number of LLP decays inside the respective detector's chamber, we take into 
consideration both the production of the mesons and hence the neutralinos, and the decay of the neutralinos 
via mesons. On the production side, since we study neutralinos produced from meson decays, we use
results published by the LHCb collaboration \cite{Aaij:2015bpa,Aaij:2016avz} for estimating $N_M$, the total 
number of the meson type ``M" produced at the LHC. For $D$-mesons, we consider only neutralinos produced 
from $D^{+}$- and $D_s$-mesons which are relevant for the benchmark scenarios we consider. 
Ref.~\cite{Aaij:2015bpa} gives the cross section for producing $D^+$, and $D^{*+}$. The latter decays to  
$D^+$-, and $D_s$-mesons at the 13-TeV LHC for a certain kinematic range: $0< p_T < 15$ GeV$/$c and 
$2.0<y<4.5$, where $p_T$ denotes transverse momentum and $y$ rapidity. We use the computer program
\texttt{FONLL} \cite{Cacciari:1998it,Cacciari:2001td,Cacciari:2012ny,Cacciari:2015fta} to extrapolate these 
numbers to the whole kinematic range, and, after taking into account the decay of the $D^{*+}$- to $D^+$-mesons, 
we obtain the total numbers of $D^+$ and $D_s$ produced over the hemisphere for $\mathcal{L}=3/$ab: 
\begin{eqnarray}
N_{D^+}&=&1.58\times 10^{16}\,, \label{eq:NDplus}\\
N_{D_s}&=&5.11\times 10^{15}\,. \label{eq:NDs}
\end{eqnarray}
At the LHC the mesons
are produced over the full $4\pi$. However the detectors we are considering here for LLPs are always off to 
one side of the collision point. We thus at first only consider the forward or backward hemisphere (2$\pi$) 
which contains the respective detector. We then later impose the necessary geometric cuts corresponding to 
a specific detector.

Among the $B$-mesons, $B^0$ and $B^\pm$ are of interest here. Ref.~\cite{Aaij:2016avz} presents the 
experimentally measured $b$-quark production cross section at the 13-TeV LHC for $2<\eta<5$, and the 
corresponding number after extrapolation over the full $\eta$ range with the numerical tool \texttt{Pythia 8} 
\cite{Sjostrand:2006za,Sjostrand:2014zea}. We take the fragmentation factors of $B$-mesons directly from 
Ref.~\cite{deVries:2015mfw}, which were obtained by simulating $1\,$M events of \texttt{HardQCD:hardbbbar} 
in \texttt{Pythia 8} \cite{Sjostrand:2006za,Sjostrand:2014zea}. We obtain 
\begin{eqnarray}
N_{B^+}&=&7.30\times 10^{14}\,, \label{eq:NBplus}\\
N_{B^0}&=&7.28\times 10^{14}\,,\label{eq:NBzero}
\end{eqnarray}
over a hemisphere for $\mathcal{L}=3/$ab. The branching ratios of these mesons decaying to neutralinos 
are easily calculated with the formulas given in Ref.~\cite{deVries:2015mfw}. We arrive at the following expression 
for the total number of neutralinos produced in a hemisphere
\begin{eqnarray}
N_{\chi}^{\text{prod}}=\sum_M N_M\cdot \text{BR}(M\rightarrow \tilde{\chi}_1^0+l),
\end{eqnarray}
where $l$ is the associated lepton in the meson decay, which can be charged or neutral.

We then apply Monte Carlo (MC) techniques to determine the average probability of the neutralinos decaying inside the 
detector chamber, 
\begin{eqnarray}
\langle P[\tilde{\chi}_1^0\text{ in d.r.}]\rangle=\frac{1}{N^{\text{MC}}_{\tilde{\chi}_1^0}}\sum_{i=1}^{N^{\text{MC}}_
{\tilde{\chi}_1^0}}P[(\tilde{\chi}_1^0)_i\text{ in d.r.}]\,,
\label{eq:decay_prob}
\end{eqnarray}
where $P[(\tilde{\chi}_1^0)_i\text{ in d.r.}]$ is the probability for a given generated neutralino to decay in the ``detectable region",
``d.r.". Dividing by the total number of simulated neutralinos produced, $N^{MC}_{\tilde\chi^0_1}$, gives the average. We  explain 
how to calculate $\langle P[\tilde{\chi}_1^0\text{ in d.r.}]\rangle$ for each detector in detail below. 

Since it is difficult to experimentally reconstruct the trajectory of the neutral final-state particles of the neutralino decays, we 
consider only charged decay products to be detectable. (See Ref.~\cite{deVries:2015mfw} for a discussion of the potential
influence of decays to $K^0$'s.) The final number of observed neutralino decays is expressed as
\begin{eqnarray}
N^{\text{obs}}_{\tilde{\chi}_1^0}=N_{\chi}^{\text{prod}}\cdot \langle P[\tilde{\chi}_1^0\text{ in d.r.}]\rangle \cdot 
\text{BR}(\tilde{\chi}_1^0\rightarrow\text{char.}).
\end{eqnarray}

We use \texttt{Pythia 8.205} \cite{Sjostrand:2006za,Sjostrand:2014zea} to perform the MC simulation in order to calculate 
$\langle P[\tilde{\chi}_1^0\text{ in d.r.}]\rangle$ in Eq.~\eqref{eq:decay_prob}. We use two matrix element calculators of 
\texttt{Pythia}, namely \texttt{HardQCD:hardccbar} and \texttt{HardQCD:hardbbbar}, to generate initial $D$- and $B$-mesons, 
respectively. Note that the differential cross section of producing heavy flavor mesons in the very forward direction, where \texttt{FASER} sits, 
is not validated in \texttt{Pythia}. In order to solve this problem, we reweigh the \texttt{Pythia} meson production cross section at different 
ranges of transverse momentum and pseudorapidity by the corresponding more reliable numbers calculated by using \texttt{FONLL}. 
We simulate 20,000 events for each benchmark scenario and extract the kinematical information of each 
neutralino $(\tilde{\chi}_1^0)_i$ from \texttt{Pythia}: $(E_i, p_i^z, \theta_i, \phi_i)$. Here the $z$-direction is along the 
beam pipe, $p_i^z$ is the component of the 3-momentum along the $z$-axis, $E_i$ is the total energy of the neutralino, 
and $\theta_i,\,\phi_i$ are the polar and azimuthal angles, respectively. With this kinematical information we derive the 
relativistic quantities as follows:
\begin{eqnarray}
\gamma_i&=&E_i/m_{\tilde{\chi}_1^0},\\
\beta_i&=&\sqrt{1-\gamma_i^{-2}},\\
\lambda_i&=&\beta_i\gamma_i/\Gamma_{\text{tot}}(\tilde{\chi}_1^0),\\
\beta_i^z&=&p_i^z/E_i, \\
\lambda_i^z&=&\beta_i^z\gamma_i/\Gamma_{\text{tot}}(\tilde{\chi}_1^0).
\end{eqnarray}
where $\Gamma_{\text{tot}}(\tilde{\chi}_1^0)$ is the total decay width of $\tilde{\chi}_1^0$ and can be calculated with 
formulas given in Ref.~\cite{deVries:2015mfw}, $\lam_i$ is the decay length of $(\tilde{\chi}_1^0)_i$ along the direction 
of its movement in the lab frame and $\lambda_i^z$ is the z-component of $\lam_i$. These quantities are used to calculate 
$P[(\tilde{\chi}_1^0)_i\text{ in d.r.}]$ for each detector. We now discuss the detectors in turn.

\subsection{CODEX-b}
\label{subsect:codex-b}

\begin{figure}[t]
\centering
  \includegraphics[width=\columnwidth]{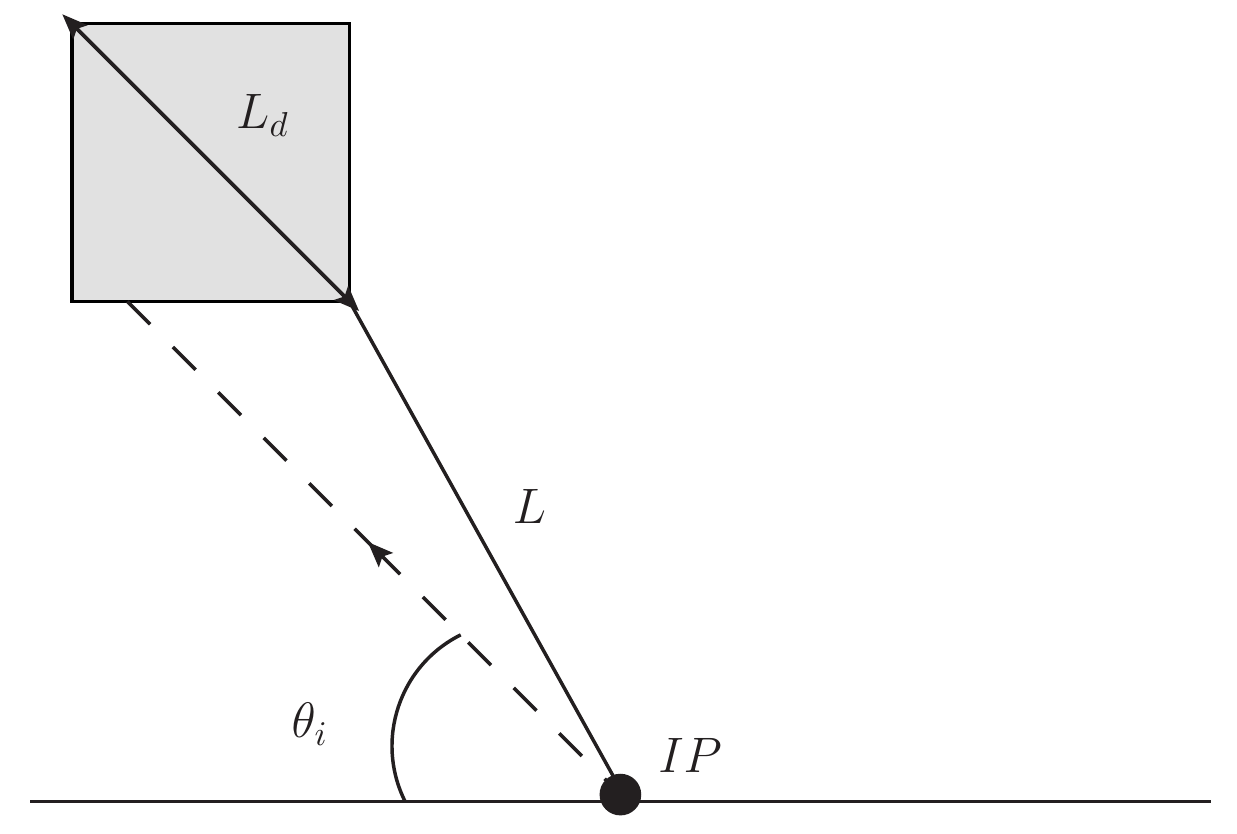}
\caption{Side-view sketch of the \texttt{CODEX-b} detector with definition of distances and angles used in text. IP denotes the 
interaction point in \texttt{LHCb}. 
The dashed line describes an example LLP track. }
\label{fig:codex-b-sketch}
\end{figure}

\texttt{CODEX-b} (``Compact detector for Exotics at LHCb") \cite{Gligorov:2017nwh} was proposed as a cubic detector with 
dimension $10^3~\text{m}^3$, sitting inside an underground cavity at a distance $L=25$ m  from the \texttt{LHCb} IP. The 
differential production distribution is flat in the azimuthal angle and the azimuthal coverage of the detector is about $0.4/2\pi
\approx6\%$. The polar angle range of the \texttt{CODEX-b} experiment at the appropriate azimuthal angle is between 
11.4$^\circ$ and 32.5$^\circ$. This corresponds to the pseudo-rapidity range $\eta\in[0.2,0.6]$. For this narrow range, 
and at the precision of this analysis, we also treat the polar angle differential production distribution as flat. As we mentioned 
earlier, \texttt{LHCb} is expected to have a total integrated luminosity of $300/$fb, smaller by one order of magnitude  than 
\texttt{ATLAS} or \texttt{CMS}. We calculate $P[(\tilde{\chi}_1^0)_i\text{ in d.r.}]$ with the following expression:
\begin{align}
\hspace{-0.31cm}P[(\tilde{\chi}_1^0)_i\text{ in d.r.}]=\left\{\begin{aligned}
&\frac{0.4}{2\pi} \cdot \frac{1-e^{-\frac{L_d}{\lambda_i}}}{e^{\frac{L}{\lambda_i}}},&& \eta_i \in[0.2,0.6],\\
&0,&&\text{~else,}
\end{aligned}
\right. \hspace{-0.5cm}
\end{align}
where we approximately treat the box detector as a spherical shell segment with the volume length $L_d=10$m. $\eta_i$ is the pseudorapidity of $(\tilde{\chi}_1^0)_i$ 
and $\eta_i = -\ln[\tan{\theta_i/2}]$. A brief sketch of the setup of \texttt{CODEX-b} is shown in Fig.~\ref{fig:codex-b-sketch}.

\subsection{FASER}
\label{subsect:faser}

\begin{figure}[ht]
\centering
  \includegraphics[width=\columnwidth]{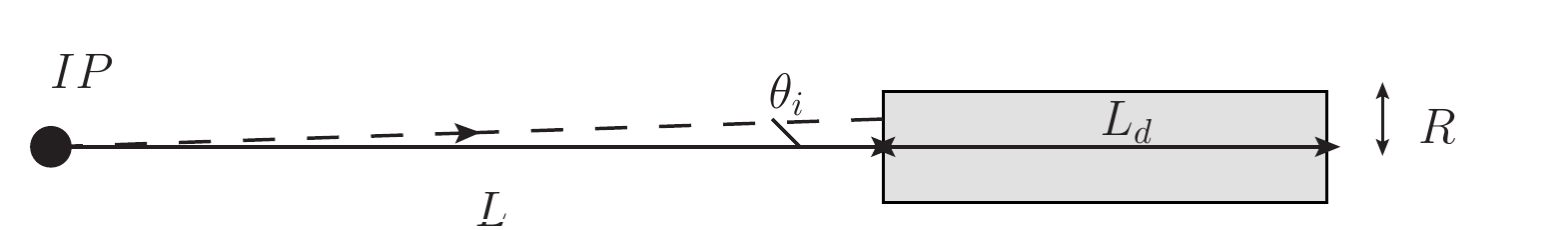}
\caption{Side-view sketch of the \texttt{FASER} detector with definition of distances and angles used in text. The dashed line 
describes an example LLP track.}
\label{fig:faser-sketch}
\end{figure}

\texttt{FASER} (``ForwArd Search ExpeRiment") \cite{Feng:2017uoz} proposes to build a small cylindrical detector placed a 
few hundred meters downstream of the \texttt{ATLAS} or \texttt{CMS} IP in the very forward region. In a series of papers 
\cite{Feng:2017uoz,Feng:2017vli,Kling:2018wct,Feng:2018pew} several different variants of \texttt{FASER} have been 
proposed. In this paper, we focus on a recent setup, which would sit at a particularly promising location in the side 
tunnel TI18 \cite{Feng:2018pew}. We denote the distance from the IP to the near end of the detector as $L=470$m, the radius 
of \texttt{FASER} as $R=1$m, and the detector length as $L_d=10$m. Following is the expression for calculating the probability 
for a given neutralino to decay inside \texttt{FASER}:
\begin{align}
P[(\tilde{\chi}_1^0&)_i\text{ in d.r.}]=\frac{1-e^{-\frac{L_i}{\lambda_i^z}}}{e^{\frac{L}{\lambda_i^z}}}\,,  \\
L_i&=\left\{\begin{aligned}
&0\,,&&\tan{\theta_i}>\frac{R}{L}\,,\\
&L_d\,,&&\tan{\theta_i}<\frac{R}{L+L_d}\,,\\
&\frac{R}{\tan{\theta_i}}-L\,,&&\text{~else\,.}
\end{aligned}
\right.
\end{align}
There is no azimuthal angle suppression because the \texttt{FASER} detector is cylindrical. Here the three cases correspond 
respectively to $1)$ the extended potential neutralino trajectory misses the decay chamber, $2)$ the extended potential 
neutralino trajectory passes through the entire length of the detector,  and $3)$ the extended neutralino trajectory exits through 
the side of the detector. In practice, we treat the third case as negligible. It corresponds to the very narrow angular range $\theta
_i\in[0.1194^\circ,0.1219^\circ]$. And furthermore the decay products of the neutralinos may exit through the side and  may thus 
miss the detector. These neutralinos hence would not be detected. A sketch of the geometric configuration of \texttt{FASER} is 
shown in Fig.~\ref{fig:faser-sketch}.

\subsection{MATHUSLA}
\label{subsect:mathusla}

\begin{figure}[ht]
\centering
  \includegraphics[width=\columnwidth]{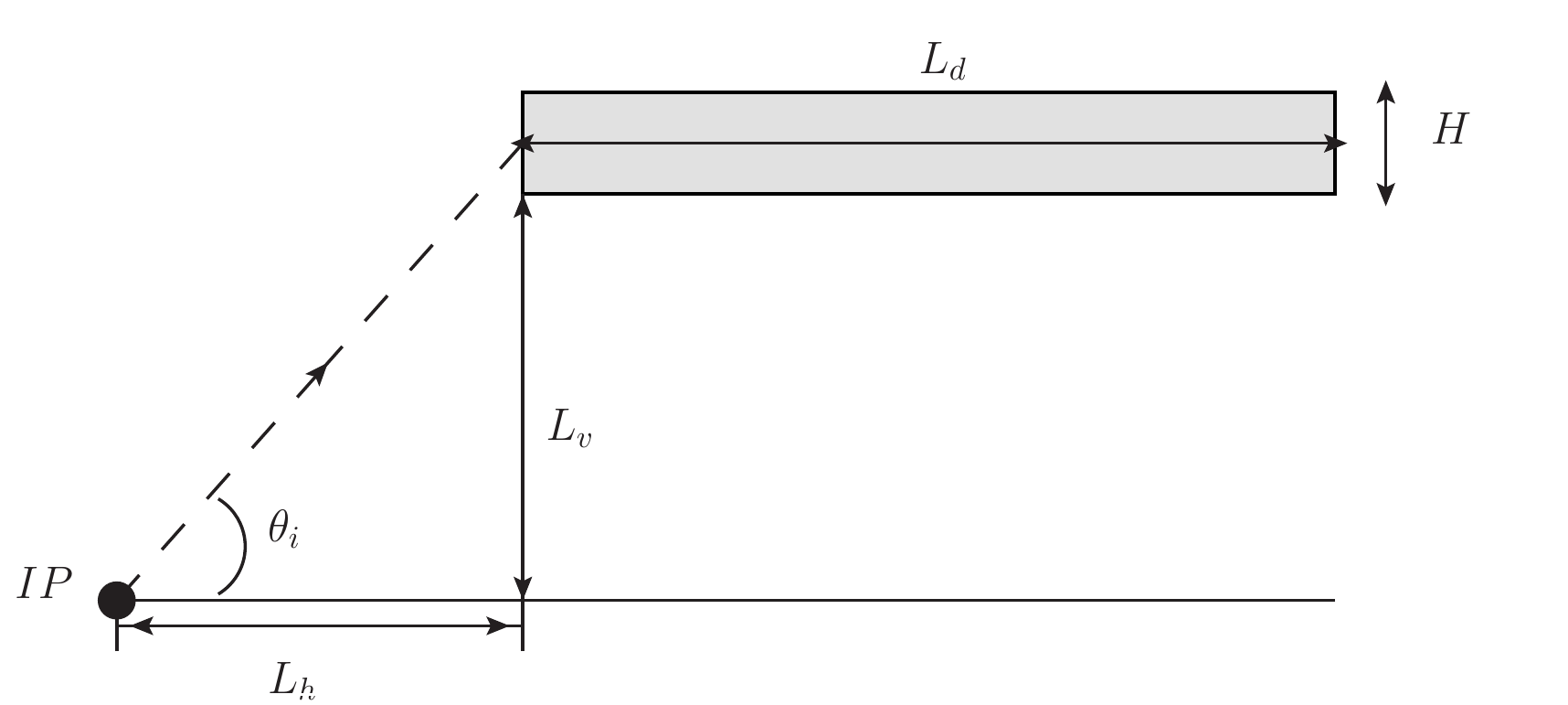}
\caption{Side-view sketch of the \texttt{MATHUSLA} detector with definition of distances and angles used in text. The dashed line describes an example LLP track.}
\label{fig:mathusla-sketch}
\end{figure}

In Ref.~\cite{Chou:2016lxi} it has been proposed to construct a surface detector $100\,$m above the \texttt{ATLAS} IP called
\texttt{MATHUSLA} (``MAssive Timing Hodoscope  for Ultra Stable neutraL pArticles"). The detector should be horizontally 
offset by $100\,$m from the \texttt{ATLAS} IP and with a massive dimension of $200$m$\times 200$m$\times20$m, 
\texttt{MATHUSLA} is expected to have excellent sensitivity for detecting LLPs. Below we show the formul{\ae} for calculating 
$P[(\tilde{\chi}_1^0)_i\text{ in d.r.}]$ in \texttt{MATHUSLA}:
\begin{align}
&P[(\tilde{\chi}_1^0)_i\text{ in d.r.}]=\frac{1}{4} \frac{1-e^{-\frac{L'_i}{\lambda_i^z}}}{e^{\frac{L_i}{\lambda_i^z}}}\,,\\
L_i&=\text{min}\bigg(\text{max}\bigg(L_h,\frac{L_v}{\tan{\theta_i}}\bigg),L_h+L_d\bigg)\,,\\
L'_i&=\text{min}\bigg(\text{max}\bigg(L_h,\frac{L_v+H}{\tan{\theta_i}}\bigg),L_h+L_d\bigg)-L_i\,.
\end{align}
Here, $L_h$ and $L_v$ are the horizontal and vertical distance from the IP to the near end of \texttt{MATHUSLA}, and they both equal $100$m. $L_d=200$m 
is the horizontal length of \texttt{MATHUSLA} and $H=20$m is its height. The factor $1/4$ comes from the azimuthal angle coverage. Both \texttt{MATHUSLA} 
and \texttt{FASER} expect to have $3/$ab luminosity of data by $\sim2035$. We show the schematic plot of 
\texttt{MATHUSLA} in Fig.~\ref{fig:mathusla-sketch}.

\newcommand{\nicefrac}[2]{#1/#2}
\section{Results}
\label{sect:results}
We present our numerical results in this section. In Ref.~\cite{deVries:2015mfw} a series of benchmark scenarios 
representative of $LQ\bar D$ couplings were investigated. In these scenarios, both the light lepton flavor (electron/muon) 
and the heavy tau flavor are considered, as the $\tau$ lepton leads to large phase space suppression effects. Also, 
different neutral or charged $D$- and $B$-mesons which would decay to the neutralino are considered; 
this is important because the cross sections of producing these mesons substantially differ, \textit{cf.} 
Eqs.~(\ref{eq:NDplus})-(\ref{eq:NBzero}).  In the present study, as a follow-up work to Ref.~\cite{deVries:2015mfw}, we 
choose to focus only on one key benchmark scenario which features the important characteristics for a comparison of the 
proposed LHC(b) detectors' sensitivities, while only briefly discussing the other scenarios. We first consider the explicit RPV 
model and then also discuss the model-independent case.

Since the operators for production and decay scale with $\nicefrac{\lam'}{m^2_{\tilde{f}}}$, we have three free parameters in 
the theory, after assuming that all SUSY fermions $\tilde f$ have degenerate masses, \textit{cf.} Sec.~\ref{setc:theory},
namely: $\nicefrac{\lam'_P}{m^2_{\tilde{f}}}$, $\nicefrac{\lam'_D}{m^2_{\tilde{f}}}$, and $m_{\tilde{\chi}_1^0}$. Here $\lam'_
{P/D}$ is the $LQ\bar D$ coupling giving rise to the production/decay of the $\tilde{\chi}_1^0$, and $m^2_{\tilde{f}}$ is the 
sfermion mass relevant for the production/decay process, respectively.\footnote{The explicit formulae including the dependence 
on the relevant sfermion masses are given in Ref.~\cite{deVries:2015mfw}.} We therefore show model-dependent plots in two 
separate planes for the aforementioned benchmark scenarios: $m_{\tilde{\chi}^0_1}$ vs. ($\nicefrac{\lambda'_P}{m^2_{\tilde
{f}}} = \nicefrac{\lambda'_D}{m^2_{\tilde{f}}}$) and $\nicefrac{\lambda'_P}{m^2_{\tilde{f}}}$ vs. $\nicefrac{\lambda'_D}{m^2_
{\tilde{f}}}$. For the latter plane, we present results for three different values of $m_{\tilde{\chi}_1^0}$.

In addition, we present model-independent results in the plane  BR vs. $c\tau$ for a generic LLP. Here $c\tau$ is the decay 
length of the LLP, and BR is the product of the branching ratios of the respective meson decaying to the LLP and of the LLP 
decaying to a charged meson and a charged lepton. These results can be interpreted in terms of any LLP which has the same 
or similar reaction chain.

For the key benchmark scenario, we choose to show all three types of plots while for the others we select only one single type 
of plot, where the distinctive features of the scenario may be best emphasized. Depending on the exact construction of the 
detectors, they can possibly also track neutral mesons. We thus show sensitivity estimates for two cases: 1)~only charged final 
states can be tracked, and 2)~both neutral and charged ones.

\subsection{Benchmark Scenario 1}
\label{subsect:scen1}

We begin with the  RPV scenario we consider in detail in this study with $D_s$-mesons produced at the LHC, which decay to a neutralino, which 
in turn travels for a macroscopic distance before decaying to a kaon and a lepton. In this scenario we assume $\lam'_{122}$ and $\lam'_{112}$ are 
the only non-vanishing $L Q \bar{D}$ couplings. $\lambda'_{122}$ gives rise to the production of $\tilde{\chi}_1^0$ via 
\begin{equation}
D_s \rightarrow \tilde{\chi}_1^0 +  e^\pm\,, \qquad  \mathrm{(production)}
\label{eq:chi-prod-via-D}
\end{equation}
and to the invisible neutralino decay 
\begin{equation}
\tilde{\chi}_1^0\to(\eta/\eta'/\phi)+\nu_e\,,\qquad \mathrm{(decay\; via\;} \lam'_{122})
\end{equation} 
On the other hand, $\lam'_{112}$ leads to both visible and invisible decays
\begin{eqnarray}
\tilde\chi^0_1\to \left\{
\begin{array}{l}
K^{(*)\pm}+e^\mp\,, \\[2.5mm]
K^0_{S,L}+\nu_e\,.
\end{array}
\right.\qquad \mathrm{(decay\;via\;} \lam'_{112})
\label{eq:vis-decay}
\end{eqnarray}
The invisible decays are important to take into account in the evaluation because they affect the total width of $\tilde{\chi}_1^0$. 
We summarize this scenario
in Tab.~\ref{tab:122-112-info}.

\begin{table}
\begin{center}
\begin{tabular}{r||l}
\hline
$\lambda^\prime_P$ for production & $\lambda^\prime_{122}$ \\
$\lambda^\prime_D$ for decay & $\lambda^\prime_{112}$ \\
produced meson(s) & $D_s$ \\
visible final state(s) & $K^{\pm} e^\mp, K^{*\pm} e^\mp$ \\
invisible final state(s) via $\lambda^\prime_P$ & $(\eta, \eta^\prime, \phi) + (\nu_e, \bar{\nu}_e)$ \\
invisible final state(s) via $\lambda^\prime_D$ & $(K^0_L,K^0_S, K^*) + (\nu_e, \bar{\nu}_e)$ \\
\hline
\end{tabular}
\caption{Features of Benchmark Scenario 1.}
\label{tab:122-112-info}
\end{center}
\end{table}

\begin{figure*}[t]
\centering
  \includegraphics[width=.45\linewidth]{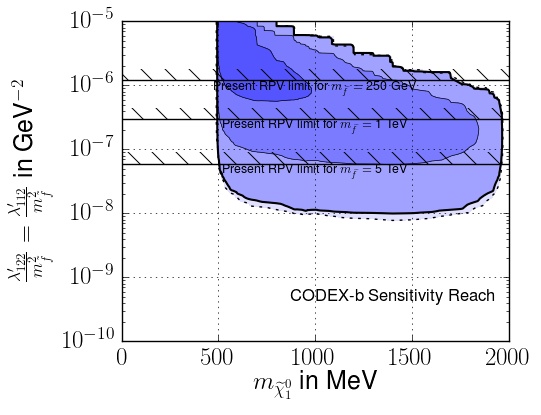}
 \includegraphics[width=.45\linewidth]{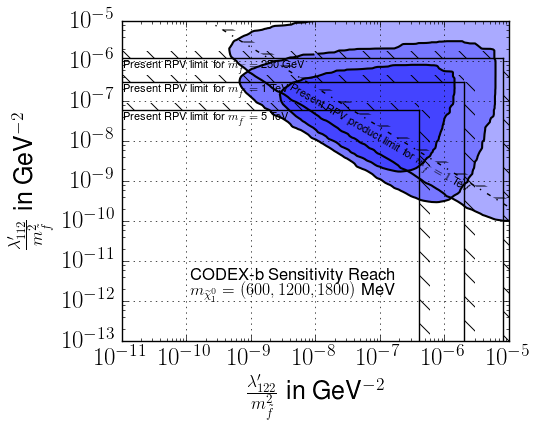}
 \\
  \includegraphics[width=.45\linewidth]{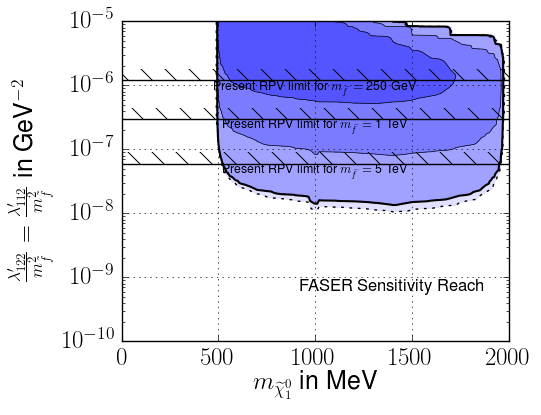}
   \includegraphics[width=.45\linewidth]{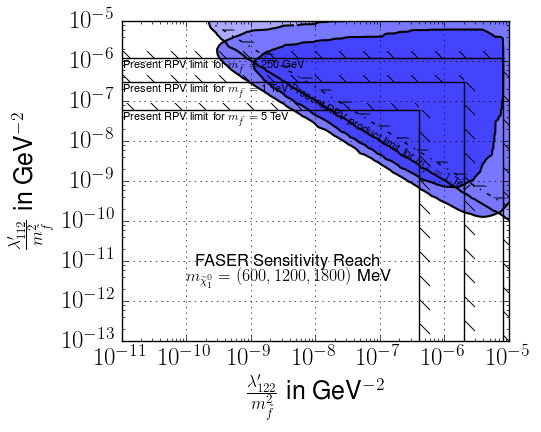}
   \\
  \includegraphics[width=.45\linewidth]{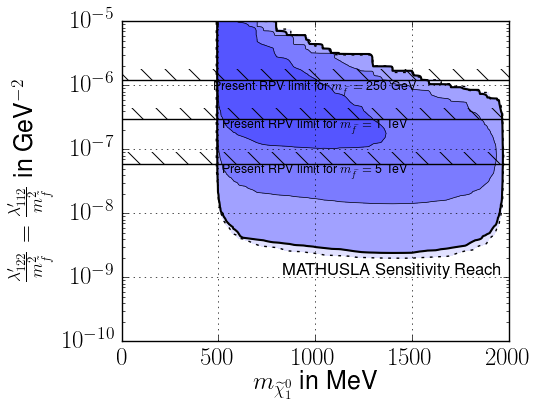}
 \includegraphics[width=.45\linewidth]{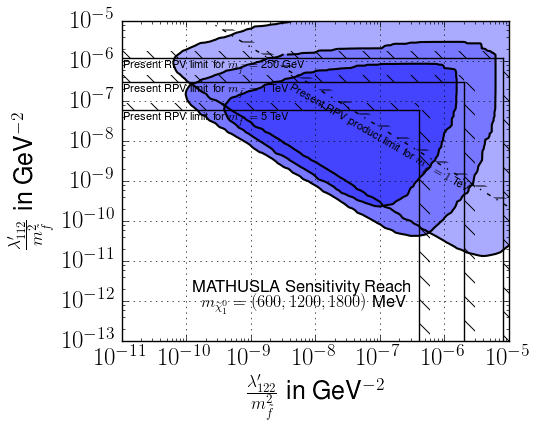}
  \\
\caption{Sensitivity estimate of \texttt{CODEX-b}, \texttt{FASER} and \texttt{MATHUSLA} for Benchmark Scenario 1. 
On the left, we show the reach in terms of $m_{\tilde{\chi}^0_1}$ and $\nicefrac{\lambda'_P}{m^2_{\tilde{f}}} = \nicefrac{
\lam'_D}{m^2_{\tilde{f}}}$. The light blue/blue/dark blue regions enclosed by the solid black lines correspond to $\geq 
3/3\times 10^3/3\times 10^6$ events. The light blue region is extended only slightly  below by a dashed curve, 
representing the extended sensitivity reach if we assume our detectors can also detect neutral decays of the neutralino. 
The hashed solid lines correspond to the single RPV couplings' limit for different sfermion masses.  On the right, the 
two couplings are not required to be identical and plots in the plane $\nicefrac{\lam'_P}{m^2_{\tilde{f}}}$ vs. $\nicefrac{\lam'_D}
{m^2_{\tilde{f}}}$ are shown for the detectors. We consider three choices of $m_{\tilde{\chi}^0_1}$: 600~MeV (light blue), 
1200~MeV (blue), 1800 MeV (dark blue). The solid hashed lines again represent the individual coupling bounds and the 
hashed dot-dashed line is the upper bound derived from the limit on the product of the two $L Q \bar D$ couplings for 
$m_{\tilde{f}}=1\,$TeV.}
\label{fig:122-112-model-dep}
\end{figure*}

We now present our results. In Fig.~\ref{fig:122-112-model-dep} we show model-dependent sensitivity estimates for the 
three detectors: \texttt{CODEX-b}, \texttt{FASER} and \texttt{MATHUSLA}. In the left column, plots are presented in the 
plane $m_{\tilde{\chi}^0_1}$ vs. ($\nicefrac{\lam'_P}{m^2_{\tilde{f}}} = \nicefrac{\lam'_D}{m^2_{\tilde{f}}}$). We have set 
$\lam'_D=\lam'_P$, and vary their values and the mass of $\tilde{\chi}^0_1$. In order to see how the number of neutralino 
decay events change with varying mass and RPV couplings, we show the light blue, blue, and dark blue areas 
corresponding to the parameter space where respectively $\geq 3$, $\geq 3\times 10^3$ and $\geq 3\times10^6$ events 
are observed. The hashed solid lines denote the present RPV limits for a set of sfermion mass values, 
Eqs.~\eqref{eq:single_boundsa}, \eqref{eq:single_boundsb}, translated to $\lam'/\tilde m^2$. We do not show the product 
bound from Eq.~\eqref{eq:product_boundsb}, as for a 1~TeV sfermion mass it coincides almost exactly with the 5 TeV 
bound on the single couplings. The bound on $\lam'/\tilde m^2$ scales linearly with the sfermion mass, when taking the 
scaling of the bound on $\lam'$ into account.

The 3-event dashed contour isocurve is extended to the lighter shaded region, bounded by a dotted line; this is obtained 
when we assume that invisible decays of the neutralinos can be detected as well. Whether this will be possible is an 
outstanding experimental question. In any case, we observe that for this benchmark scenario this would only give a very 
small extension in the sensitivity reach.

The range of sensitivity in the neutralino mass $m_{\tilde\chi^0_1}$ is strictly determined by the kinematics of 
the production and decay
\begin{equation}
(M_{K^\pm}+m_e)<m_{\tilde\chi^0_1} < (M_{D_s}-m_e)\,.
\end{equation}
and is thus identical for the three experiments. The range in sensitivity in $\lam'/\tilde m^2$ is determined by the 
experimental set-up. Comparing the results for the three detectors, we find that for this model  \texttt{CODEX-b} and 
\texttt{FASER} reach similar values of $\nicefrac{\lambda'}{m^2_{\tilde{f}}}$, while \texttt{MATHUSLA} is more sensitive by
a factor $\sim 5$. Furthermore they can all extend well beyond existing low-energy limits on the R-parity violating 
couplings.

On the right in Fig.~\ref{fig:122-112-model-dep}, we show plots in the plane $\nicefrac{\lambda'_P}{m^2_{\tilde{f}}}$ vs. 
$\nicefrac{\lambda'_D}{m^2_{\tilde{f}}}$ for three values of $m_{\tilde{\chi}^0_1}: 600\,$MeV (light blue region)$, 1200$ 
MeV (blue region)$, 1800$ MeV (dark blue region). In this benchmark scenario, $\lambda'_P=\lambda'_{122}$ and 
$\lam'_D=\lambda'_{112}$. For these results, the requirement that $\lambda'_P=\lambda'_D$ is lifted, so we observe an 
interplay between the production and decay of $\tilde{\chi}^0_1$. We may compare each detector's sensitivity range in 
different parameters. For example, the $\nicefrac{\lambda'_P}{m^2_{\tilde{f}}}$ reach of \texttt{FASER} is only weaker 
than that of \texttt{MATHUSLA} by a factor $\sim 3$, even though \texttt{FASER} is more than 25,000 times smaller than 
\texttt{MATHUSLA}. This arises because \texttt{FASER} exploits very well the advantage of receiving the light $D$-mesons 
(and the produced neutralinos) boosted 
in the very forward direction, where the differential production cross section is significantly higher. As for the reach in $\nicefrac{\lambda'_D}{m^2_{\tilde{f}}}$, 
\texttt{MATHUSLA} shows again the strongest potential. Here we include single coupling bounds as solid hashed lines for three different sfermion masses (250,
1000 and 5000 GeV) and now also the product bound as a dashed hashed line for a 1 TeV sfermion mass. Again all experiments are sensitive well beyond 
existing limits.

\afterpage{\FloatBarrier}
\begin{figure}[t]
\centering
  \includegraphics[width=\columnwidth]{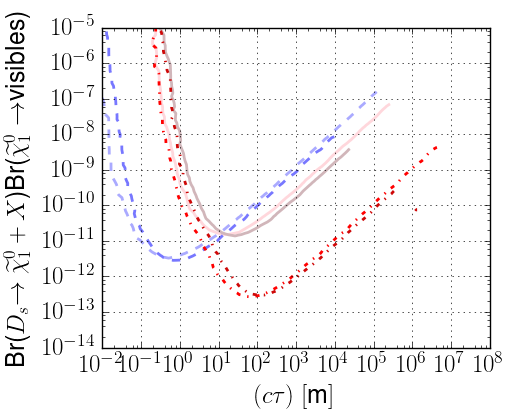}
\caption{Model-independent sensitivity estimate of \texttt{CODEX-b} (solid pink curves), \texttt{FASER} (dashed blue curves) and \texttt{MATHUSLA} 
(dot-dashed red curves) for Benchmark Scenario 1. We show the sensitivity reach as isocurves of 3 events of visible decays. 
For the axes, we choose the neutralino's unboosted decay length $c \tau$ and the relevant meson branching ratio.   The 
light/medium colors correspond respectively to a 600/1200 MeV neutralino mass. $\lam'_P= \lam'_{122},~\lambda'_D=\lambda'_{112}.$}
\label{fig:122-112-model-indep}
\end{figure}

We next consider a model-independent description, where we interpret our results in terms of the physical observables, BR$=\mathrm{BR}_P\cdot\mathrm
{BR}_D$, instead of the RPV-SUSY parameters. Here 
\begin{eqnarray}
\mathrm{BR}_P&=&\mathrm{BR}(D_s\to\mathrm{LLP}+e^\pm)\,, \\
\mathrm{BR}_D&=&\mathrm{BR}(\mathrm{LLP}\to K^{(*)\pm} + e^\mp)\,,
\end{eqnarray} 
\noindent and we allow for any LLP. The results are shown in  Fig.~\ref{fig:122-112-model-indep}. The dashed blue isocurves 
are for 3 events of visible decays inside the \texttt{FASER} decay chamber for the two lighter LLPmass values $m_
{LLP}$ values of those considered in Fig.~\ref{fig:122-112-model-dep}. The solid pink curves are for \texttt{CODEX-b} 
and the dot-dashed red for \texttt{MATHUSLA}. The light and medium colors correspond respectively to the smallest 
(600 MeV) and medium (1200 MeV) values for the LLP mass we choose to investigate. We do not show the curve for 
the heaviest mass value (1800 MeV) because it is almost the same as that for 1200 MeV. The $c\tau$ position of the 
valley of the isocurves, the point of maximal sensitivity, is determined by
\begin{equation}
\langle \beta\gamma\rangle c\tau \approx \langle L\rangle\,,
\end{equation}
where $\langle \beta\gamma\rangle$ is the average boost of the neutralinos flying in the direction of the detector and $\langle 
L\rangle$ is the distance from the IP to the middle of the respective detector. We estimate $\langle \beta\gamma\rangle$ of 
the neutralinos that fly inside each detector by simulating 10,000 events in each case, and summarize the results for each 
Benchmark Scenario and detector in Tab.~\ref{tab:mean_betagamma}. The values for $\langle L\rangle$ are
\begin{align}
\langle L\rangle = \left\{ \begin{aligned} &30.0~ \mathrm{m}&& \mathrm{for}\ \mathtt{\CODEXb}, \\
&475~\mathrm{m} &&\mathrm{for}\  \mathtt{FASER}, \\
&223~\mathrm{m}&&\mathrm{for}\  \mathtt{MATHUSLA} 
\end{aligned} \right.
\end{align}
Using the values of Benchmark Scenario 1, we get for the most sensitive $c\tau$ value
\begin{align}
(c\tau)_{\mathrm{max.\;sensitivity}} = \left\{ \begin{aligned} &18.3~ \mathrm{m}&& \mathrm{for}\ \mathtt{\CODEXb}, \\
&0.85~\mathrm{m} &&\mathrm{for}\  \mathtt{FASER}, \\
&77~\mathrm{m} &&\mathrm{for}\  \mathtt{MATHUSLA} 
\end{aligned} \right.
\end{align}
which agrees with Fig.~\ref{fig:122-112-model-indep}.

\begin{table}[t]
\begin{center}
\begin{tabular}{l|c c c c }
\hline
Benchm. Sc. & $m_{\tilde{\chi}_1^0}$ (MeV) & $\langle \beta\gamma\rangle_{\texttt{CODEX-b} }$ &  $\langle \beta\gamma\rangle_{\texttt{FASER}}$ &  $\langle \beta\gamma\rangle_{\texttt{MATHUSLA}}$ \\
 \hline
1 ($D_s$) & 1200 & 1.64 & 560 & 2.87 \\
 2 ($D^\pm$) & 1200 & 1.50 & 682 & 2.90 \\
 3 ($B^0$ \& $\bar{B}^0$)  & 1000 & 4.07 & 793 & 7.32 \\
 4 ($B^0$ \& $\bar{B}^0$) & 2000 & 2.22 & 391 & 3.88 \\
 5 ($B^0$ \& $\bar{B}^0$) & 2500 & 1.88 & 308 & 3.36 \\
 5 ($B^\pm$) & 2500 & 1.55 & 358 & 2.95 \\
\hline

\end{tabular}
\caption{Summary of $\langle \beta\gamma\rangle$ values for each detector in all the Benchmark Scenarios. Inside the 
parenthesis in each column, the type of the mother meson of the neutralino is given. In particular, in Benchmark Scenario 
5, neutralinos can be produced from decay of either $B^0$ or $B^\pm$; therefore we show the results in two 
separate rows. }
\label{tab:mean_betagamma}
\end{center}
\end{table}

The BR position of the valleys is  determined by the luminosity of the experiment, the cross section of producing $D_s$-mesons, 
the pseudorapidity coverage, the volume of the detector and the product of the branching ratios. The BR reach of \texttt{CODEX-b} 
is roughly one order of magnitude larger than that of \texttt{FASER}. This is mainly due to the fact that \texttt{LHCb} has a one 
order of magnitude lower projected luminosity than that of \texttt{ATLAS}/\texttt{CMS}. Perhaps more importantly, in spite of the 
huge volume difference between \texttt{MATHUSLA} and \texttt{CODEX-b}/\texttt{FASER}, the BR reach in \texttt{MATHUSLA} is 
only one order of magnitude stronger than that in \texttt{FASER}. For large $c\tau$ values \texttt{MATHUSLA} performs far better 
than \texttt{CODEX-b}, but for shorter neutralino lifetimes the detectors perform equally well. The reason is that the distance traveled 
to \texttt{MATHUSLA} is about ten times larger than for \texttt{CODEX-b}, such that less neutralinos reach the former detector for 
short-lived neutralinos. This leads to a similar sensitivity despite the larger integrated luminosity and the larger detector size of 
\texttt{MATHUSLA}.

Note that Fig.~\ref{fig:122-112-model-indep} is very similar to the first plot of Fig.~1 in Ref.~\cite{Helo:2018qej}, the result of which 
was obtained in the context of a Type-I Seesaw model, where the right-handed neutrino is the LLP with a mass of 1 GeV produced 
from $D$-meson decays. This illustrates the model-independence of the results shown in the BR-$c\tau$-plane.

\subsection{Benchmark Scenario 2}
\label{subsect:scen2}
Now we briefly study the other the benchmark scenarios. In Benchmark Scenario 2, $\lam'_P=\lam'_{121}$  instead 
of $\lam'_{122}$, so that a $D^{\pm}$, instead of a $D_s$, decays to the lightest neutralino. Correspondingly, the 
invisible final states due to $\lam'_P$ are now kaons, instead of $\eta,\,\eta',\,\phi$. The relevant information is summarized 
in Tab.~\ref{tab:121-112-info}.

\begin{table}[t]
\begin{center}
\begin{tabular}{r||l}
\hline
$\lambda^\prime_P$ for production & $\lambda^\prime_{121}$ \\
$\lambda^\prime_D$ for decay & $\lambda^\prime_{112}$ \\
produced meson(s) & $D^\pm$ \\
visible final state(s) & $K^{\pm} e^\mp$, $K^{*\pm} e^\mp$ \\
invisible final state(s) via $\lambda^\prime_P$ & $(K^0_L,K^0_S, K^*) + (\nu_e, \bar{\nu}_e)$ \\
invisible final state(s) via $\lambda^\prime_D$ & $(K^0_L,K^0_S, K^*) + (\nu_e, \bar{\nu}_e)$ \\
\hline
\end{tabular}
\caption{Features of Benchmark Scenario 2.}
\label{tab:121-112-info}
\end{center}
\end{table}

\begin{figure}[b]
\centering
  \includegraphics[width=\columnwidth]{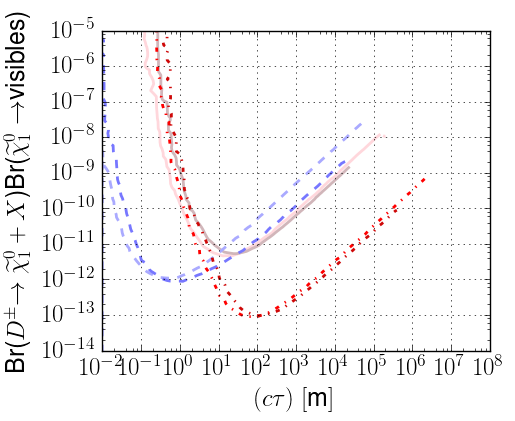}
  \\
\caption{Model-independent sensitivity estimate of \texttt{CODEX-b}, \texttt{FASER} and \texttt{MATHUSLA} for Benchmark 
Scenario 2. The format is the same as in Fig.~\ref{fig:122-112-model-indep}. The plot corresponds to visible decay products 
only. The two LLP mass values are 600 and 1200 MeV. $\lambda'_P=\lambda'_{121},~\lambda'_D=\lambda'_{112}.$}
\label{fig:121-112-model-indep}
\end{figure}

The model-dependent results are very similar to those shown in Fig.~\ref{fig:122-112-model-dep}. We do not show them 
again. One difference is that the low-energy product bound is stricter in this case, \textit{cf.} Eq.~(\ref{eq:product_boundsa}). 
It is due to $K^0\!-\!\bar K^0$-mixing and scales linearly with the \textit{sneutrino} mass. As pointed out in 
Ref.~\cite{deVries:2015mfw}, in the case of \texttt{SHiP}, also here, if the sneutrino mass is equal to the relevant squark 
mass of production and decay, the sensitivity reach of \texttt{CODEX-b} and \texttt{FASER} is excluded by these low-energy 
bounds. If there is a strong hierarchy and the sneutrinos are (unexpectedly) significantly heavier than the relevant squarks, 
this scenario is still viable. All the same, we present the model-independent results in Fig.~\ref{fig:121-112-model-indep} for 
the same LLP mass values as in Benchmark Scenario 1: 600 and 1200 MeV.  We again drop the curve for the 1800~MeV 
neutralino mass. The main difference between Fig.~\ref{fig:121-112-model-indep} and Fig.~\ref{fig:122-112-model-indep} is 
the BR reach. Benchmark Scenario 1 has a weaker BR reach mainly because $N_{D^+}\simeq 3\cdot N_{D_s}$, \textit{cf.} 
Eqs.~(\ref{eq:NDplus}), (\ref{eq:NDs}), and the neutralinos have a smaller branching ratio to charged particles.

\begin{table}[t]
\begin{center}
\begin{tabular}{r||l}
\hline
$\lambda^\prime_P$ for production & $\lambda^\prime_{131}$ \\
$\lambda^\prime_D$ for decay & $\lambda^\prime_{112}$ \\
produced meson(s) & $B^0, \bar{B}^0$ \\
visible final state(s) & $K^{\pm} e^\mp, K^{*\pm} e^\mp$ \\
invisible final state(s) via $\lambda^\prime_P$ & none \\
invisible final state(s) via $\lambda^\prime_D$ & $(K^0_L,K^0_S, K^*) + (\nu_e, \bar{\nu}_e)$ \\
\hline
\end{tabular}
\caption{Features of Benchmark Scenario 3}
\label{tab:131-112-info}
\end{center}
\end{table}
\subsection{Benchmark Scenario 3}
\label{subsect:scen3}

\begin{figure}[t]
\centering
  \includegraphics[width=\columnwidth]{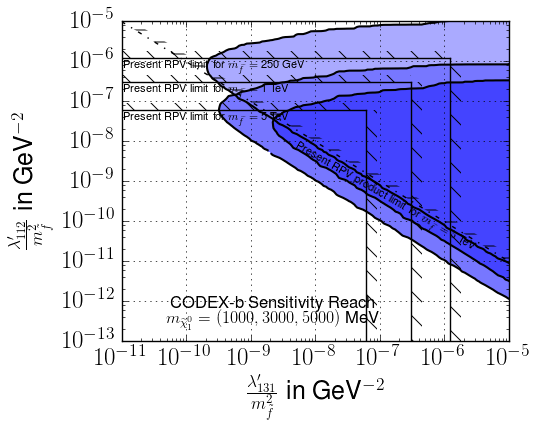}
   \\
  \includegraphics[width=\columnwidth]{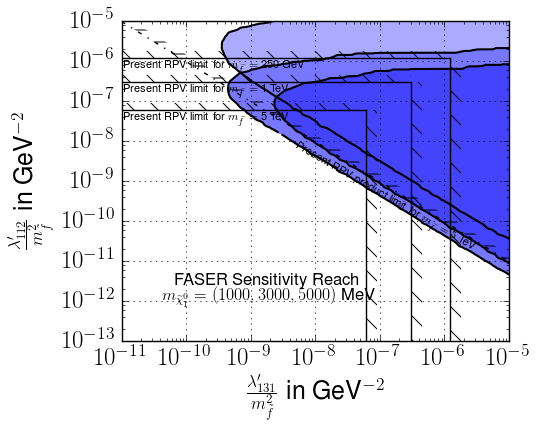}
  \includegraphics[width=\columnwidth]{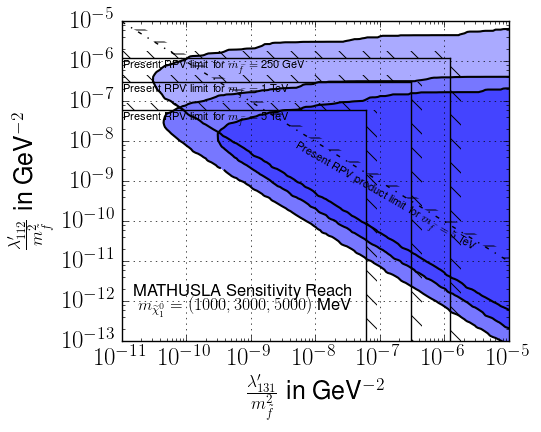}
  \\
\caption{Model-dependent sensitivity estimate of \texttt{CODEX-b}, \texttt{FASER} and \texttt{MATHUSLA} for Benchmark 
Scenario 3. The format is 
the same as in Fig.~\ref{fig:122-112-model-dep}.} 
\label{fig:131-112-model-dep}
\end{figure}

We now study several scenarios where  bottom mesons decay to a neutralino. Since the bottom mesons are much heavier than the charm mesons, 
the mass reach improves compared to the previous scenarios. In the present Benchmark Scenario 3, as before, we have $\lam'_D=\lam'_{112}$ 
giving  both invisible and visible neutralino decays. For the neutralino production we have $\lam'_P=\lam'_{131}$ such that $B^0$ (and $\bar{B}^0$) 
decay to a neutralino. This is summarized in Tab.~\ref{tab:131-112-info}. Kinematically we can thus probe 
\begin{equation}
(M_{K^\pm}+m_e)<m_{\tilde\chi^0_1} < (M_{B^0}-m_e)\,.
\end{equation}

For this scenario we only show the model-dependent plots in the plane $\nicefrac{\lam'_P}{m^2_{\tilde{f}}}$ vs. 
$\nicefrac{\lambda'_D}{m^2_{\tilde{f}}}$ in Fig.~\ref{fig:131-112-model-dep}. Here the sensitivity reach in $\nicefrac
{\lam'_{P/D}}{m^2_{\tilde{f}}}$ is only slightly weaker than the previous $D$-meson scenarios, mainly because at 
the LHC, the cross section of producing $B$-mesons is only smaller than that of $D$-mesons by a factor of $\sim 10-50$, 
\textit{cf.} Eqs.~(\ref{eq:NDplus})-(\ref{eq:NBzero}). In the previous scenarios, \texttt{CODEX-b} shows similar sensitivity 
reach in $\nicefrac{\lam'_{P/D}}{m^2_{\tilde{f}}}$ to that of \texttt{FASER}, but now the former exceeds the latter, despite 
the fact that its projected luminosity is smaller by one order of magnitude. This is because the $B$-meson mass is more 
than twice the $D$-meson mass, and hence the produced $B$-mesons are not as much boosted in the very forward 
direction as the $D$-mesons. For the same reason, we also have a larger sensitive mass range than in the previous 
benchmark scenarios. \texttt{MATHUSLA} again has the most extensive sensitivity range.

\subsection{Benchmark Scenario 4}
\label{subsect:scen4}

\begin{figure}[t]
\centering
  \includegraphics[width=\columnwidth]{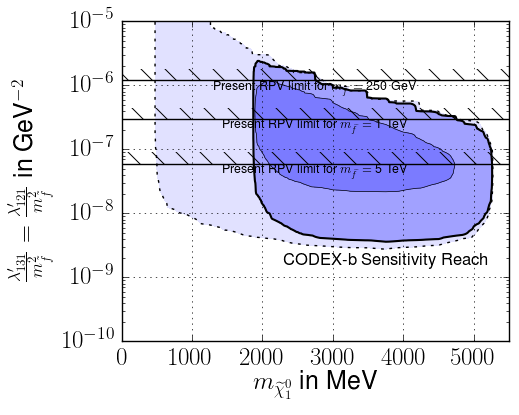}
 \\
  \includegraphics[width=\columnwidth]{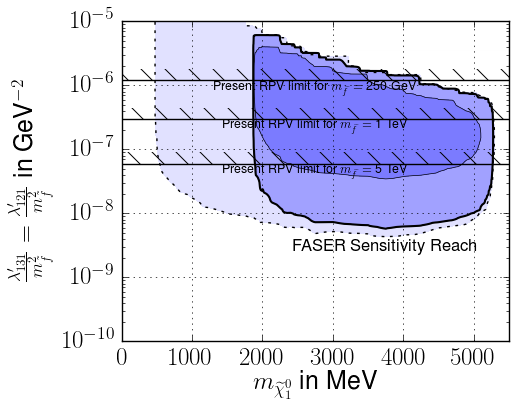}
  \includegraphics[width=\columnwidth]{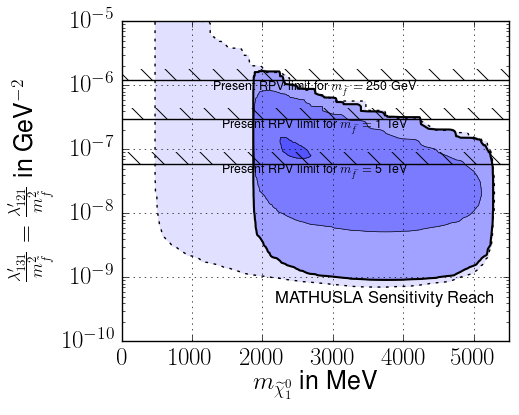}
  \\
\caption{Model-dependent sensitivity estimate of \texttt{CODEX-b}, \texttt{FASER} and \texttt{MATHUSLA} for Benchmark Scenario 4. The format is the same as in Fig.~\ref{fig:122-112-model-dep}.}
\label{fig:131-121-model-dep}
\end{figure}

In this scenario, we use the same $\lambda'_P$ as in the previous scenario, but change $\lam'_D$ from 
$\lam'_{112}$ to $\lambda'_{121}$. Correspondingly the decay mode of the neutralino changes from the
decay to a $K^\pm$ to a $D^{\pm}$, though the invisible decay mode remains the same. We 
summarize the relevant information in Tab.~\ref{tab:131-121-info}. The kinematic reach in the neutralino mass is
\begin{equation}
(M_{D^\pm}+m_e)<m_{\tilde\chi^0_1} < (M_{B^0}-m_e)\,,
\end{equation}
for the charged decay modes. It is extended when the invisible modes are included by replacing $M_{D^\pm}\to M_{K^0}$.
\begin{table}[t]
\begin{center}
\begin{tabular}{r||l}
\hline
$\lambda^\prime_P$ for production & $\lambda^\prime_{131}$ \\
$\lambda^\prime_D$ for decay & $\lambda^\prime_{121}$ \\
produced meson(s) & $B^0, \bar{B}^0$ \\
visible final state(s) & $D^{\pm} e^\mp, D^{*\pm} e^\mp$ \\
invisible final state(s) via $\lambda^\prime_P$ & none \\
invisible final state(s) via $\lambda^\prime_D$ & $(K^0_L,K^0_S, K^*) + (\nu_e, \bar{\nu}_e)$ \\
\hline
\end{tabular}
\caption{Features of Benchmark Scenario 4}
\label{tab:131-121-info}
\end{center}
\end{table}
We present the results of this scenario in the plane $(\nicefrac{\lambda'_P}{m^2_{\tilde{f}}} = \nicefrac{\lam'_D}
{m^2_{\tilde{f}}})$ vs. $m_{\tilde{\chi}^0_1}$ in Fig.~\ref{fig:131-121-model-dep}. The lower mass sensitivity is 
now raised up to the $D$-meson mass, if only the visible decays of the neutralinos are considered. If we 
consider the detectors able to track neutral final states, the lower mass sensitivity is dramatically extended, 
as expected, down to $m_K\sim 500$ MeV. For large values of $(\nicefrac{\lam'_P}{m^2_{\tilde{f}}} = \nicefrac{\lam'_
D}{m^2_{\tilde{f}}})$ we produce many more neutralinos, but they now mostly decay before reaching the 
detector. That is why there is no sensitivity here. For very small values of $(\nicefrac{\lam'_P}{m^2_{\tilde{f}}} = 
\nicefrac{\lam'_D}{m^2_{\tilde{f}}})$ we produce too few neutralinos and the neutralinos decay well after the 
detector. Otherwise  the results are  similar to those in Benchmark Scenario 3.

\afterpage{\FloatBarrier}
\newpage
\subsection{Benchmark Scenario 5}
\label{subsect:scen5}
While the previous benchmark scenarios concern only the light electron, we here explore the effect of the heaviest lepton 
$\tau$ on the sensitivity estimates. We consider $\lam'_P=\lam'_{313}$ and $\lam'_D=\lam'_{312}$. Then both $B^0 (\bar{B}
^0)$ and $B^\pm$ may decay to a neutralino. In particular, $B^\pm$ decays then include a $\tau$ lepton along with a 
neutralino. This gives a large suppression in phase space and a correspondingly lower neutralino mass sensitivity. While 
$\lam'_P$ does not induce any invisible decay of the lightest neutralino, $\lam'_D=\lam'_{312}$ leads to both visible decays 
to a kaon and a $\tau$, and invisible decays to a kaon and a $\nu$.  We summarize the information in Tab.~\ref{tab:313-312-info}. 
The mass sensitivity is given by
\begin{align}
m_{\tilde\chi^0_1}  &> 
\left\{
\begin{array}{cl}
M_{K^\pm}+m_\tau\;&\mathrm{if\, only\, obs.\, visible\, decays,} \\
M_{K^0}\;&\mathrm{if\, also\, obs.\ invisible\, decays} \\
\end{array}
\right.\ , \\
m_{\tilde\chi^0_1} &< 
\left\{\begin{array}{cl}
M_{B^\pm}-m_\tau\;&\mathrm{if\, produced \,via}\,B^\pm\\
M_{B^0}\;&\mathrm{if\, produced \,via}\,B^0\\
\end{array}
\right. \ .
\end{align}

\begin{table}[t]
\begin{center}
\begin{tabular}{r||l}
\hline
$\lambda^\prime_P$ for production & $\lambda^\prime_{313}$ \\
$\lambda^\prime_D$ for decay & $\lambda^\prime_{312}$ \\
produced meson(s) & $B^0, \bar{B}^0, B^\pm (+\,\tau^\mp)$  \\
visible final state(s) & $K^{\pm} \tau^\mp, K^{*\pm} \tau^\mp$ \\
invisible final state(s) via $\lambda^\prime_P$ & none \\[1mm]
invisible final state(s) via $\lambda^\prime_D$ & $(K^0_L,K^0_S, K^*) + (\nu, \bar{\nu})$ \\
\hline
\end{tabular}
\caption{Features of Benchmark Scenario 5. At the end of the third row we emphasize that the charged 
$B$-meson decay to the neutralino is accompanied by a tau lepton.}
\label{tab:313-312-info}
\end{center}
\end{table}

\begin{figure}[t]
\centering
  \includegraphics[width=\columnwidth]{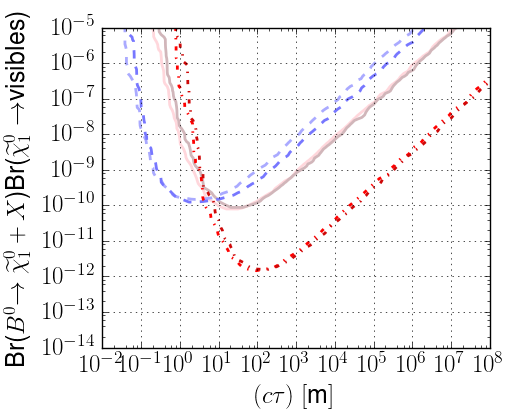}
\\
\caption{Model-independent sensitivity estimate of \texttt{CODEX-b}, \texttt{FASER} and \texttt{MATHUSLA} 
for the $B^0$-case of Benchmark Scenario 5. The format is the same as in Fig.~\ref{fig:121-112-model-indep}. The two mass values are 2500 and 3750 MeV. $\lam'_P=\lam'_{313},~\lam'_D=\lam'_
{312}.$}
\label{fig:313-312-model-indep}
\end{figure}

We present a plot for this benchmark scenario in the plane BR vs. $c\tau$ in Fig.~\ref{fig:313-312-model-indep}. 
We restrict ourselves to the production via $B^0$. As can be seen in Tab.~\ref{tab:mean_betagamma}, between the $B^0$ 
and the $B^\pm$ cases the $\langle\beta\gamma\rangle$ values, which determine the $c\tau$ sensitivity, differ 
between 15 and 20\%. This is below the resolution of our logarithmic plot. For the BR sensitivity the dominant 
contribution is the $B^0$ vs. $B^\pm$ production rate, however they are almost identical, \textit{cf.} 
Eqs.~(\ref{eq:NBplus}), (\ref{eq:NBzero}). The only real difference is that for $B^\pm$ we must have $m_{\tilde
\chi^0_1}\lsim 3500\,$MeV, \textit{i.e.} the medium mass case (3750~MeV) is not possible.

In Fig.~\ref{fig:313-312-model-indep} the labeling is similar to the previous scenarios. For each detector, 
\textit{i.e.} \texttt{FASER} (blue), \texttt{CODEX-b} (pink) and \texttt{MATHUSLA} (red), the light/medium colors 
correspond to the lightest(2500~MeV)/medium(3750~MeV) $m_{\tilde{\chi}_1^0}$. 
Fig.~\ref{fig:313-312-model-indep} is very similar to the right panel of Fig.~1 of Ref.~\cite{Helo:2018qej} where 
the sensitivity to sterile neutrinos was discussed. This again shows that most features of the figures are relativity insensitive to 
the nature of the LLP.

\afterpage{\FloatBarrier}
\subsection{Decay Branching Ratios of the $\tilde{\chi}_1^0$}
\begin{figure}[t]
\centering
  \includegraphics[width=\columnwidth]{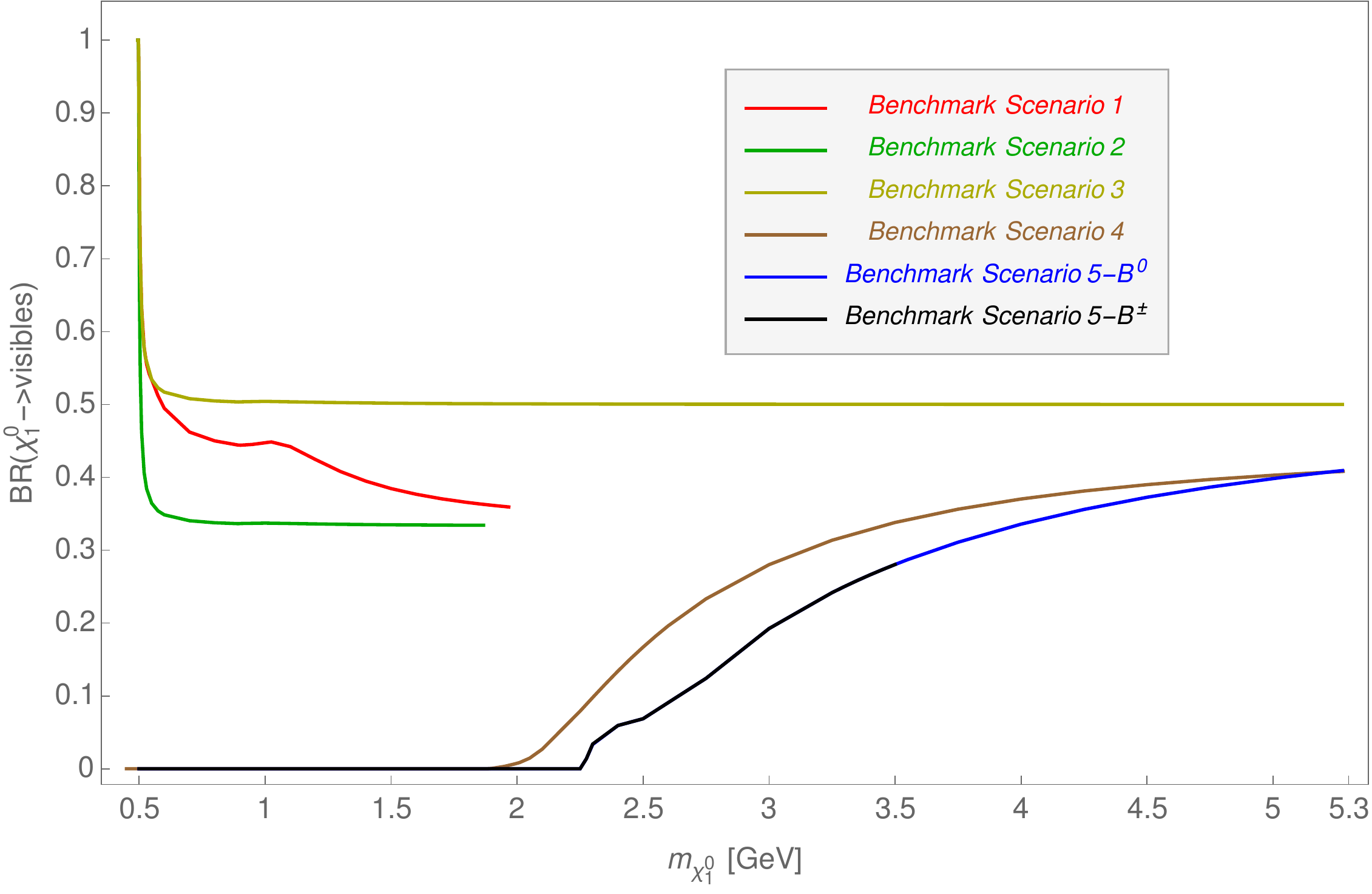}
\\
\caption{Branching ratios of $\tilde{\chi}_1^0$ to visible states in Benchmark Scenarios 1-5 as a function of $m_{\tilde{\chi}_1^0}$ [GeV], where 
we set $\lam'_P=\lam'_D$. For each curve, only the kinematically allowed range of $m_{\tilde{\chi}_1^0}$ is plotted.
}
\label{fig:BRofneutralino}
\end{figure}
After having presented results of different benchmark scenarios in the previous subsections, we supplement our results by showing in Fig.~\ref{fig:BRofneutralino}
the decay branching ratios of the $\tilde{\chi}_1^0$ to visible, \textit{i.e.} charged meson final states,  as a function of $m_{\tilde{\chi}_1^0}$ in the 
kinematically allowed range for all the scenarios. The curves can be well understood by considering the kinematic thresholds for the various neutralino
decays. For Benchmark Scenarios 1, 2 and 3 the first decay channel to open is to the charged kaon: $\tilde\chi^0_1\to K^\pm e^\mp$. Thus the visible branching 
ratio starts at 1. It rapidly drops as the $K^0$-threshold is crossed. The asymptotic value of the branching ratios in Benchmark Scenarios 1, 2 and 3 is simply
determined by the number of charged or neutral decay channels. For example, in Benchmark Scenario 2 above all thresholds there are 4 visible final
states and 8 invisible final states, giving a branching ratio to visible of 4/(4+8)=1/3. The bump in Benchmark Scenario 1 (red curve) is due to the extra threshold of 
the $\eta$-meson below the $K^{+*}$ and $K^{0*}$ masses. Note that it is the vector meson $K^{0*}$ which is relevant for the neutralino decay, not the 
pseudoscalar \cite{deVries:2015mfw}. 

In Benchmark Scenarios 4 and 5 the first kinematically accessible final state is invisible: $K^0\nu_e$. Thus for low values of the neutralino
mass the visible branching ratio vanishes. In Benchmark Scenario 4 the first visible decay mode to open up is: $D^+e^-$, in Benchmark Scenario 5 it
is: $K^+\tau^-$. We also point out that we divide Scenario 5 into two cases: the neutral bottom meson decaying to a neutralino, and the 
charged bottom meson decaying to a neutralino accompanied by a tau lepton. This leads to the overlap of the two corresponding curves, the blue and the black 
ones, below a mass of $\sim 3.5$ GeV. Above that mass value, only $B^0$ may decay to such a neutralino, because of the large mass of the tau lepton.

\afterpage{\FloatBarrier}
\subsection{Summary of our Results}

We finish this section with some conclusions drawn from the above results for specific benchmarks. For all scenarios, 
we observe similar reach in $\nicefrac{\lambda'}{m^2_{\tilde{f}}}$ for the two experiments  \texttt{CODEX-b} and 
\texttt{FASER} and the strongest sensitivity for \texttt{MATHUSLA}. Even though \texttt{FASER} takes good advantage 
of the boost of the $D$- and $B$-mesons in the very forward direction, \texttt{MATHUSLA} overcomes this disadvantage 
by virtue of its much larger volume. Compared to earlier results determined for the \texttt{SHiP} experiment \cite{deVries:2015mfw}, 
both \texttt{FASER} and \texttt{CODEX-b} have a smaller expected reach in $\nicefrac{\lambda'}{m^2_{\tilde{f}}}$. Even 
\texttt{MATHUSLA} cannot outperform \texttt{SHiP} in scenarios with $D$-meson dependence, because \texttt{SHiP}'s 
centre-of-mass energy of $\approx$ 27~GeV results in very high sensitivity. For models with $B$-meson decays, 
however, we expect \texttt{MATHUSLA}'s sensitivity to be comparable or even better than \texttt{SHiP}s.

We also translated our results into sensitivity limits on the meson's branching ratio BR and here we observe differences 
in the experimental sensitivities of \texttt{FASER} and \texttt{CODEX-b}, depending on the meson flavour: while for the $B$-meson's BR we observe a similar 
ranking of the two experiments in the reach for $\nicefrac{\lambda'}{m^2_{\tilde{f}}}$, \texttt{FASER}'s reach is expected 
to be slightly stronger than \texttt{CODEX-b}'s in case of scenarios with $D$-meson decays. In all cases \texttt{MATHUSLA} 
shows the largest sensitivity reach.

As discussed above, our results in terms of $c \tau$ vs. BR can be used to estimate experimental sensitivities for long-lived 
particles different from RPV neutralinos. We used our results for Benchmark Scenario 1 as an example by pointing out the 
strong resemblance of Fig.~\ref{fig:122-112-model-indep} and the first plot of Fig.~1 in Ref.~\cite{Helo:2018qej} determined 
for a different BSM model albeit with similar decay topology. Though we use this observation to label these limits as 
\textit{model independent}, our combined set of results still points to several sources of model-dependence. An important 
degree of freedom to mention here is the LLP mass. For fixed  $c \tau$ and branching ratio, changing the mass of the LLP 
has an important impact on its kinematic parameters $\beta \gamma$ and $\eta$ which, as we have shown, play a non-negligible 
role as they are strongly connected to the preferred distance  $L$ of the detector to the primary interaction vertex. Another 
important aspect neither covered in $c \tau$ nor the LLP BR is the overall topology which leads to the production of the LLP. 
Though all our benchmarks share the same topology $p p \rightarrow \text{meson} + X, \text{meson} \rightarrow \text{LLP} 
+ Y$, we observe sizable differences in the experimental coverages depending on the flavor of the produced meson, here 
$D$ or $B$. Not only do these have different total production cross sections but also their own kinematics --- and with that 
the kinematics of the LLPs they decay into --- differ.

A full ``model-independent'' analysis would require the consideration of several additional degrees of freedom, some of 
which cannot be formulated as a continuous parameter like the overall production-and-decay-topology of the LLP. This 
results in an unfeasible, if not impossible, exercise. Nevertheless, although the dependence on these additional parameters 
may not be explicitly covered in our chosen degrees of freedom $c \tau$ and BR, our results can still be applied to a large 
class of LLP models different from RPV as long as they share similarities to the topologies discussed here.

\section{Conclusions}
\label{sect:conclusion}
We have investigated the sensitivity of three recently proposed detectors at the LHC: \texttt{CODEX-b}, \texttt{FASER}, and 
\texttt{MATHUSLA} with respect to the detection of light long-lived neutralinos in RPV-SUSY scenarios. The neutralinos are 
produced and decay via the RPV $L Q \bar D$ operator with coupling $\lam'$. We studied five representative benchmark 
scenarios of the RPV couplings proposed in Ref.~\cite{deVries:2015mfw} where a similar sensitivity study for \texttt{SHiP} 
was completed. In general \texttt{CODEX-b} and \texttt{FASER} show similar reach in $\nicefrac{\lambda'}{m^2_{\tilde{f}}}$, 
where $m_{\tilde f}$ is the mass of supersymmetric fermion partners, while \texttt{MATHUSLA} performs better by approximately 
one order of magnitude. Comparing \texttt{MATHUSLA} results with \texttt{SHiP} estimates, we find that \texttt{MATHUSLA} 
shows a better sensitivity in scenarios involving B-meson while it provides only comparable or slightly weaker results than 
\texttt{SHiP} for models in which neutralinos interact with $D$-mesons.

We also want to point out that cosmic rays may provide an additional argument to choose an underground experiment like 
\texttt{FASER} over a surface experiment like \texttt{MATHUSLA}. Although we ignored this allegedly controllable background 
contamination in our analysis, the required workload to fully control this background source may be significantly different between 
the experiments discussed here.

\bigskip
\centerline{\bf Acknowledgements}

\bigskip
H.K.D. and Z.S.W. are supported by the Sino-German DFG grant SFB CRC 110 ``Symmetries and the Emergence of Structure in QCD". D.D. acknowledges funding and support from DFG grant SFB 676, project B1. We thank 
Oliver Freyermuth and Peter Wienemann for kind help with using the department cluster. We also thank Felix Kling for useful discussions on FASER. 
H.K.D. would like to thank the Galileo Galilei Institute for Theoretical Physics for the hospitality and the INFN for partial support during the completion 
o this work.

\bigskip


\end{document}